# A numerical attempt on the Chiral Schwinger Model

T.D. Kieu [a] [b]  and P. S. Hawkins [c]

[a]School of Physics, University of Melbourne, Parkville Vic 3052, Australia

[b]School of Natural Sciences, Institute for Advanced Study, Princeton NJ 08540, USA

[c]Computer Centre, Monash University, Clayton Vic 3168, Australia

The chiral Schwinger model is formulated in the wilson-fermion formulation on the lattice and then simulated by the complex langevin algorithm. The simulation is done both without and with gauge fixing to the Lorentz gauge for the compact gauge links. Some preliminary results are presented which indicate that the complex langevin is well behaving with the complex chiral fermion determinant.

## 1. Introduction

In this firstever attempt, in our knowledge, to simulate a chiral gauge theory on the lattice we will work with the simplest abelian theory in two dimensions, the chiral Schwinger model (CSM) [1]. Even though apparently anomalous, the theory is consistent. In general, we have some indications that so-called anomalous chiral gauge theory is consistent even in four dimensions and with non-abelian gauge groups [2, 3].

The standard pure-gauge wilson action for compact gauge links is adopted in our work. To deal with the fermion doubling we employ wilson fermions for the lattice fermion action as in [4]. In addition to the left-handed fermion that couples directly to the gauge links we also have the right-handed part necessitated by the wilson term. However, we do not include the gauge links in the wilson term in order to preserve the Goltermann-Petcher shift symmetry [5] in the massless limit. The symmetry entails the decoupling of the right-handed part.

The complex-valuedness of the chiral fermion determinant is an obstacle in the computer simulation of theory of this type. Here we enlist the complex langevin algorithm for the task despite the fact that the algorithm cannot shown to converge to the right weight in general. However, it is expected to work well by the simple argument of analytic continuation if the imaginary part of the determinant is relatively small. And we will justify the algorithm *posterior*.

From the exact solution of CSM it can be seen that the determinant imaginary part will vanish in the Lorentz gauge if there is gauge invariance to legitimise the gauge fixing. We have argued elsewhere that, despite of its appearance, the CSM derives its consistency from a gauge-invariant theory containing a Wess-Zumino-type term in the gauge fields [2]. More than a coincidence, in the Lorentz gauge the WZ term vanishes with the determinant imaginary part since the two cancel each other to keep the theory anomaly-free. We will exploit this fact to the full and simulate the theory with Lorentz gauge fixing. On the other hand, simulation is also done without gauge fixing for comparison and for conformity with the more conventional treatment of the CSM.

## 2. Complex langevin algorithm

Details of langevin simulation in lattice gauge theory with dynamical fermions can be found in reference [6]. We adopt the Runge-Kutta formulation here and calculate the chiral fermion contribution to the langevin equation by the stochastic estimate, with a variation of the conjugate gra-dient [7] for the inversion of the fermion matrix. We update the random complex vectors of the stochastic estimate twice for each update of the noise term in the langevin equation since there are two distinct steps in the Runge-Kutta method. The noise term is taken to be gaussian and real, thus the complex-valuedness is solely due to the appearance of chiral fermions.

## 3. Stochastic gauge fixing

The CSM is not gauge-invariant without the WZ term but this non-local term has not been built in into our lattice action due to the difficulty with its non-locality. To deal with this difficulty we follow every langevin step with a gauge-fixing step to keep the WZ term vanishingly small at all time. The chiral determinant imaginary part is consequently small, which would help the correct convergence of the complex langevin. The data presented in the next section are all obtained from the cold start, a special case of the Lorentz gauge condition. The gauge links of subsequent langevin times are stochastically gauge fixed to the Lorentz gauge. Such gauge fixing is a generalisation of Zwanziger procedure for non-compact fields and is not integrable [8]. The advantage for this procedure, it is claimed, is that the Gribov problem associated with covariant gauge can be avoided.

We also simulate the CSM without any gauge fixing to investigate the consistency of the model in the anomalous formulation if the complex langevin converges correctly in this case.

## 4. Preliminary results and remarks

We present some preliminary results on $32 \times 32$ lattice without and with chiral fermions, at wilson $\kappa = 0.2$. Periodic boundary conditions are imposed on both the gauge links and the fermions with the langevin step size chosen to be 0.001. For each set of parameters, 140 configurations are collected from the cold start with 400 intervening steps. This seems sufficient for thermalisation, at least for the average plaquette.

When the gauge is fixed, a value of the gauge fixing parameter is chosen large enough to keep the gauge fields confined in the Gribov horizon, but not too large so as the whole procedure is destabilised. The square of the lattice Lorentz condition is plotted versus langevin time in figure 1. The spikes there represent the occasional departures from the gauge surface.

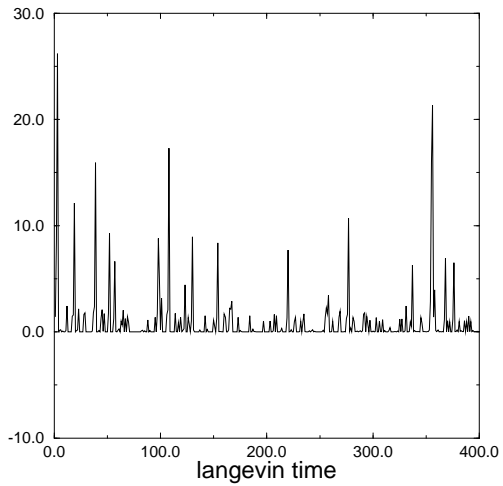

Figure 1. Lorentz gauge fixing.

We plot the average plaquette, represented by the filled and unfilled circles, versus $\beta$ in figure 2. In both cases of without and with fermions there is no sign of any phase transition. The product of the average plaquette with $\beta$ is also represented by filled and unfilled triangles in the same plot. A straight dotted line is drawn to guide the eyes in what is possibly the scaling region starting from $\beta \approx 1.2$; the data at $\beta = 3.0$ seem to suffer from the finite size effects. With few preliminary data at hand, we do not see any difference between the two cases of without and with gauge fixing for gauge-invariant expectation.

We have also considered the zero momentum gauge propagators in the lorentz gauge for various values of $\beta$. As expected, the photons have the tendency to be massive in the presence of chiral gauge interactions. More detailed analysis will be needed to extract the mass value in the continuum limit.

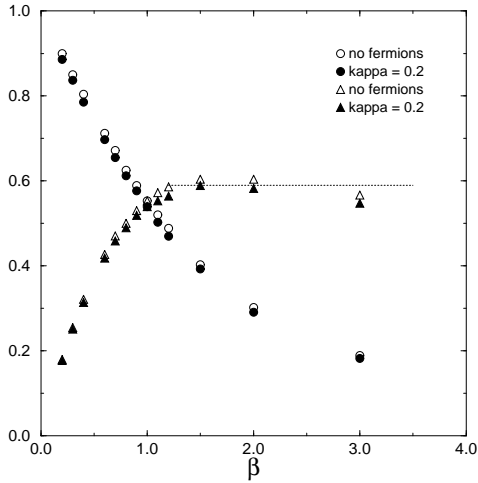

Figure 2. The average plaquette, filled and unfilled circles, and average plaquette times $\beta$, filled and unfilled triangles, versus $\beta$.

Detailed description and results of this work will be presented elsewhere. Work is also in progress to explore the phase space in $\beta$ and $\kappa$ but a suitable order parameter is yet to be defined. The complex langevin procedure apparently behaves very well for the CSM even without gauge fixing. The model provides a fertile testing ground for the complex langevin. The model, being gauge invariant in appropriate formulation, also provides an opportunity to realise the Roma approach of BRST fine tunning with extra counterterms [9] when the chiral gauge symmetry is explicitly broken by regularisation. With this in mind, we have not included the gauge links in the wilson term in the hope to minimise the number of possible counterterms in the fine tunning.

**Acknowlegement**

The simulation is being done on the DECmpp 1200Sx MP-2. PSH thanks Mitch Mason for technical support. TDK wishes to thank Claudio Parrinello and Brian Pendleton for discussions and support. He also acknowledges the hospitality of Steve Adler and the Institute for Advanced Study during a working visit, and the support of the Australian Research Council.